\newlength{\extralineskip}
\documentclass[12pt]{article}

\begin{document}
\begin{titlepage}
\begin{flushright}
          \begin{minipage}[t]{12em}
          \large UAB--FT--440\\
                 March 1998
          \end{minipage}
\end{flushright}
\vspace{\fill}

\vspace{\fill}

\begin{center}
\baselineskip=2.5em

{\large \bf Astrophysical constraints on superlight gravitinos}\footnote{Invited talk
at the 5th Workshop on High Energy Physics Phenomenology(WHEPP-5), Pune, India, 12-26
January 1998.}
\end{center}

\vspace{\fill}

\begin{center}
{\bf J.A. Grifols}\\
\vspace{0.4cm}
     {\em Grup de F\'\i sica Te\`orica and Institut de F\'\i sica
     d'Altes Energies\\
     Universitat Aut\`onoma de Barcelona\\
     08193 Bellaterra, Barcelona, Spain}
\end{center}
\vspace{\fill}

\begin{center}
\large Abstract
\end{center}
\begin{center}
\begin{minipage}[t]{36em}
I review the constraints on the mass of gravitinos that follow from
considerations on energy loss in stars and from Big Bang Nucleosynthesis
arguments.
\end{minipage}
\end{center}

\vspace{\fill}

\end{titlepage}

\clearpage

\addtolength{\baselineskip}{\extralineskip}

{\bf 1. STELLAR ENERGY LOSS\cite{raffelt}

1a. The Sun}
 
 Thermal stars like our sun are systems where thermal pressure balances
gravity. An important feature of these stars is their negative specific heat.
They heat up when the system looses energy. Indeed, when the star gives off
energy, its internal plus gravitational energy goes down and as a cosequence
the star contracts. This is so because the gravitational energy itself
diminishes, as required by the virial theorem. Contraction, in turn, implies a
rise in temperature and extra nuclear burning because the internal energy has
to compensate for the decrease of gravitational energy (virial theorem,
again). Now, the added nuclear energy supply leads to an opposite cycle of
expansion and cooling (negative specific heat) and hence to a lesser nuclear
burning. This is a self-regulating mechanism where equilibrium is maintained
by balancing the energy loss with nuclear fusion.

 Weakly interacting particles, if not trapped in the interior of stars, drain
energy. Hence they lead to accelerated consumption of nuclear fuel. Models of
stellar evolution tolerate non-standard exotic energy drain mechanisms as long
as their associated luminosity does not exceed the solar luminosity, i.e.

\begin{equation} \label{luminosity}
L \leq 10^{33} erg/s
\end{equation}

{\bf 1b. Degenerate Stars}

 These are systems dominated by the Fermi pressure of electrons (white dwarfs)
or of nucleons (neutron stars). In this case the internal energy (Fermi
energy) is nearly independent of the temperature and a loss of energy is at
the expense of gravitational contraction (which increases the the internal
Fermi energy but not the temperature). In fact the system actually cools.

 Here also, new hypothetical particles provide for extra energy depletion
which should not exceed
\begin {equation} \label{lumwd}
5\times10^{-3} erg/g/s
\end{equation}
for white dwarfs.

{\bf 1c. Type II supernovae}

 Very massive stars (much heavier than the sun) ignite all nuclear fuels up to
iron. Beyond this point no more nuclear energy is available. When the iron
core reaches its Chandrasekhar limit, the star enters an unstable
catastrophical regime: gravitational collapse follows.

{\it Gravitational collapse}

 After the thermonuclear cycle is completed, Si having transformed into Fe,
collapse is triggered by the photodissociation of $Fe^{56}$ and electron
capture on nuclei. The core loses Fermi pressure from the electrons
which no longer hold gravitational pressure and the core implodes.
Hence, $\nu_{e}$ emission dominates during infall. As the core collapses
temperature and density rise. For densities above $2\times10^{11}
g/cm^{3}$ neutrinos become trapped in the core. When the inner core
reaches nuclear densities ($\sim3\times10^{14}g/cm^{3}$) a nuclear phase
transition occurs, i.e. bound nuclei become free nucleons. This has a dramatic
effect. The core stiffens because now non-relativistic nucleons dominate the
pressure (before, the pressure of relativistic electrons and electron
neutrinos could not halt collapse). The inner core bounces generating a shock
wave which is responsible for the ejection of the mantle of the star. 

 The hydrodynamical core collapse happens in less than a second. The relevant
dynamic time scales are on the order of 10 to 100 miliseconds (almost
free-fall). For instance, the initial electron-neutrino burst lasts for about
5 ms. Finally, the proto-neutron star originated that way
cools off to a cold neutron star in a time scale of several seconds.

Independently of the details of collapse, to form a neutron star 
\begin{equation}\label{binener}
E_{B}\sim G M^{2}/R\sim 3\times10^{53} erg
\end{equation}
have beeen released. The total luminosity in electromagnetic radiation plus
kinetic energy of the ejecta in a supernova explosion is $\leq10^{51} erg$.
Emission in gravitational radiation is at most $ 1\%$. THE BULK OF THE BINDING
ENERGY ($\geq99\%$) IS EMITTED IN FORM OF NEUTRINOS. In the neutronization
burst $ \sim10^{52} erg $ (10\% of the total energy) are emitted. The rest of
the energy is radiated essentially as thermal neutrinos. 

 Observational data from Supernova 1987A (IMB, Kamiokande)\cite{hirata} imply
\begin{equation}\label{obs}
E_{\nu}\geq2\times10^{53}erg
\end{equation}
emitted over a diffusion period of the order of ten seconds. Comparing
equations (4) and (3) we see that any additional energy drain is allowed
whenever 
\begin{equation}\label{lumsn}
L_{X}\leq10^{52}erg/s
\end{equation}

{\bf 2. A CASE STUDY: BOUNDS ON GRAVITINO MASSES}

{\it The Gravitino}

  The gravitino is the spin 3/2 partner of the graviton. It is a Majorana
particle with only transverse degrees of freedom before supersymmetry
breakdown. After eating the goldstino, it  acquires the longitudinal degrees of
freedom, i.e. the 
$\pm$1/2 helicities. In some recent models the gravitino can be
superlight. Indeed, models where gauge interactions mediate the breakdown of
supersymmetry
\cite{dine}, models where an anomalous U(1) gauge symmetry induces SUSY breaking
\cite{bine}, and no-scale models are all examples of models where a superlight gravitino
can be accommodated \cite{ellis}. In all of them, the gravitino is the LSP and,
furthermore, its couplings to matter and radiation are inversely proportional to its
mass. For very small gravitino masses, the longitudinal components of the
gravitino dominate the interactions with matter so that effectively,
\begin{equation}\label{goldst}
G_{\mu}\cong\sqrt{2\over3} m_{3/2}^{-1}\partial_{\mu}\chi
\end{equation}
 
 In a subclass of the models mentioned above, the scalar and pseudoscalar
partners of the goldstino are also ultralight. Their couplings do also show the
enhancing $m_{3/2}^{-1}$ factor\cite{roy}.

{\bf 2a. Astrophysics}\cite{nowa}

{\it Supernova constraints}\cite{moha,grif,toldra,dicus}

 According to Luty and Ponton\cite{lute}, photons have an effective coupling to 
gravitinos with the following structure
\begin{equation}\label{lupo}
\delta
L={e\over2}(M/\Lambda^{2})^{2}\partial^\mu\chi\sigma^\nu\bar\chi
F_{\mu\nu}+h.c.
\end{equation}
where M is a model dependent supersymmetric mass parameter and $\Lambda$ is the
supersymmetry breaking scale. This interaction provides the leading
contribution to gravitino pair emissivity via electron-positron annihilation,
nucleon-nucleon bremsstrahlung, plasmon decay, and photon-electron scattering
followed by radiation of the gravitino pair.

 The energy-loss rate (per unit volume) via $ pn\rightarrow pn
\tilde{G}\tilde{G} $ is,
\begin{equation} \label{brems}
Q=\int  {d^3k_1\over(2\pi)^3 2k_1^0}{d^3k_2\over(2\pi)^3
2k_2^0}\prod_{i=1}^4{d^3p_i\over(2\pi)^3 2p_i^0}f_1f_2(1-f_3)(1-f_4)(2\pi)^4
\delta^4(P_f-P_i)\sum_{spins} \mid M_{fi} \mid^2(k_1^0+k_2^0)
\end{equation}
where $(p^0,\vec{p})_i$ are the 4-momenta of the initial and final state nucleons,
$(k^0,\vec{k})_{1,2}$ are the 4-momenta of the gravitinos and $ f_{1,2}$ are the
Fermi-Dirac distribution functions for the initial proton and neutron and $(1-f_{3,4})$
are the final Pauli blocking factors for the final proton and neutron. The squared matrix
element can be factorised as follows,
\begin{equation} \label{factor}
\sum_{spins} \mid M_{fi} \mid ^2=(2\pi)^2\alpha^2(M/\Lambda^2)^4N_{\mu\nu}
G_{3/2}^{\mu\nu}
\end{equation}
where $N_{\mu\nu}$ is the nuclear (OPE) tensor and $G_{3/2}^{\mu\nu}$ is the gravitino
tensor in the matrix element squared. The factor $N_{\mu\nu}$ is common to any
bremsstrahlung process involving nucleons. It appears, e.g., in neutrino bremsstrahlung
calculations and in axion bremsstrahlung calculations, and is given explicitly in ref
\cite{raffelt}. On the other hand, $G_{3/2}^{\mu\nu}$ is a tensor specific to gravitino
bremsstrahlung. It reads,
\begin{equation} \label{gravten}
G_{3/2}^{\mu\nu}=k_{1}^\mu k_{2}^\nu+k_{2}^\mu k_{1}^\nu-k_{1}.k_{2}g^{\mu\nu}
\end{equation}
The integration of $N_{\mu\nu}$ over the phase-space of the nucleons can be performed
explicitly and the details can be found in Raffelt's book \cite{raffelt}.
When we contract the result with the gravitino tensor $G_{3/2}$ and perform
the integrals over gravitino momenta to complete the energy depletion rate, we
are led to the following emissivity:
\begin{equation} \label{emiun}
Q_{brems}^{ND}
=(8192/385\pi^{3/2})\alpha^2\alpha_{\pi}^2(M/\Lambda^2)^4Y_{e}n^2_BT^{11/2}/m^{5/2}_p
\end{equation}
for non-degenerate and non-relativistic nucleons ($\alpha_\pi$ is the pionic
fine-structure constant, $n_B$ is the number density of baryons, and $Y_e$ is the mass
fraction of protons). However, nucleons are moderately degenerate in the SN core. The
emissivity in the (extreme) degenerate case is calculated to be,
\begin{equation} \label{emidos}
Q_{brems}^D=(164\pi^3/4725)\alpha^2\alpha_\pi^2(M/\Lambda^2)^4p_FT^8
\end{equation}
with $p_F$, the Fermi momentum of the nucleons. Numerically, for the actual conditions of
the star, both emissivities differ by less than an order of magnitude (about a factor of
three). Since the actual emissivity interpolates between these two values, we shall adopt
the smallest of the two (i.e. $Q^{ND}_{brems}$) to make our (conservative) estimates. We
turn next to the annihilation process.

 The energy loss for the process $e^+(p_1)+e^-(p_2) \rightarrow
\tilde{G}(k_1)+\tilde{G}(k_2)$ can be calculated along similar lines as above. The spin
averaged matrix element squared is in this case,
\begin{equation} \label{matrix}
\sum_{spins}\mid
M_{fi}\mid^2=(2\pi)^2\alpha^2(M/\Lambda^2)^4E_{\mu\nu}(p_1,p_2)G_{3/2}^{\mu\nu}(k_1,k_2)
\end{equation}
where $E_{\mu\nu}(p_1,p_2)$ equals formally the tensor $G_{3/2}^{\mu\nu}$ in eq.(10)
with
$k_1,k_2$ replaced by $p_1,p_2$.
The luminosity then is found to be,
\begin{equation} \label{lum}
Q_{ann}=8\alpha^2(M/\Lambda^2)^4T^4e^{-\mu/T}\mu^5b(\mu/T)/15\pi^3
\end{equation}
with $b(y)\equiv(5/6)e^yy^{-5}(F_5^+ F_4^- +F_4^+F_5^-)$
where 
$F_{m}^\pm(y)=\int_0^\infty dxx^{m-1}/(1+e^{x\pm y})$ ($\mu$ is the chemical
potential of the electrons). The function
$b(y)\rightarrow1$ in the degenerate limit.
 Finally, our estimate of the plasmon decay luminosity is,
\begin{equation} \label{plasm}
Q_P=16\zeta(3)\alpha^4T^3\mu^6(M/\Lambda^2)^4/81\pi^5
\end{equation}
(where only transverse plasmons have been taken into account).

The process $\gamma e\rightarrow e\tilde{G} \tilde{G} $ has not been evaluated
analitically but numerically has been seen to be of the same order as
$e^{+}e^{-}\rightarrow\tilde{G}\tilde{G}$\cite{dicus}. Taken at face value, the
bremsstrahlung rate is the largest of the four. However,
$Q_{brems}$ is overestimated since we did not consider multiple scattering
effects which are present in a dense medium\cite{raffelt}. Indeed, as for the axion
case\cite{raffelt}, the gravitino bremsstrahlung rate probably saturates around $20\%$
nuclear density and this should be taken into account when evaluating eq.(6). If we use
now the values
$T=50 MeV$,
$\mu=300 MeV$, and
$Y_e=0.3$, eqs. (6) (with $n_B\sim 0.2 n_{nuc}$), (9) and (10) give 
\begin{equation} \label{ratios}
Q_{ann}:Q_{brems}:Q_P\approx   1.2\times10^3:3\times10^2:1
\end{equation}

Therefore, a limit on $\Lambda$
will follow from the requirement that $L_{3/2} \approx VQ_{ann}$ ($V$ is the volume of
the stellar core) should not exceed $10^{52}ergs/s$. This constraint on the gravitino
luminosity
$L_{3/2}$ implies, in turn,
\begin{equation} \label{bound}
\Lambda\geq300 GeV (M/43 GeV)^{1/2}(T/50 MeV)^{11/16}(R_c/10 Km)^{3/8}
\end{equation}
or, using  $m_{3/2}=2.5\times 10^{-4}eV(\Lambda/1TeV)^2$,
\begin{equation} \label{mass2}
m_{3/2}\geq 2.3 \times 10^{-5}eV.
\end{equation}

Of course, the previous calculation makes sense only if gravitinos, once produced,
stream freely out of the star without rescattering. That they actually do so, for
$\Lambda\geq300 GeV$, can be easily checked by considering their mean-free-path in the
core. The main source of opacity for gravitinos is the elastic scattering off the Coulomb
field of the protons:
\begin{equation} \label{mfp}
\lambda=1/\sigma n=(4/\pi\alpha^2)Y_e^{-1}\rho^{-1}m_p^{-1}(\Lambda^2/M)^4
\end{equation}
 The thermally averaged cross-section for elastic gravitino scattering on electrons is
roughly a factor $T\mu/m_p^2$ smaller than that on protons and thus it does not
contribute appreciably to the opacity.
 Putting numbers in eq.(19) we find:
\begin{equation} \label{mfp2}
\lambda \simeq1.4\times10^7 cm(43 GeV/M)^4(\Lambda/300 GeV)^8
\end{equation}

 On the other hand, the calculation of Q breaks down for $\lambda\leq10Km$, i.e. for
$\Lambda\leq220GeV$, when gravitinos are trapped in the SN core. In this case,
gravitinos diffuse out of the dense stellar interior and are thermally radiated from a
gravitino-sphere $R_{3/2}$. Because in this instance the luminosity is proportional to
$T^4$, only for a sufficiently large $R_{3/2}$(where the temperature is correspondingly
lower), the emitted power will fall again below the nominal $10^{52}erg/s$. Consequently,
gravitino emission will be energetically possible, if $\Lambda$ is small enough.
 The gravitino-sphere radius can be computed from the requirement that the optical depth
\begin{equation} \label{depth}
\tau=\int_R^\infty dr/\lambda(r)
\end{equation}
be  equal to $2/3$ at $R=R_{3/2}$. Here, $\lambda(r)$ is given in eq.(19) with the
density profile ansatz:
\begin{equation} \label{rho}
\rho(r)=\rho_c(R_c/r)^m
\end{equation}
with $\rho_c=8\times10^{14}g/cm^3$, $R_c=10Km$ and $m=5-7$ and which satisfactorily
parameterises the basic properties of SN1987A \cite{turner}.
 An explicit calculation renders:
\begin{equation} \label{sphere}
R_{3/2}=R_c[(8Y_e/3\pi\alpha^2)(\Lambda^2/M)^4(m-1)/\rho_cR_cm_p]^{1/1-m}
\end{equation}
 Stefan-Boltzmann's law implies for the ratio of gravitino to neutrino luminosities,
\begin{equation} \label{SB}
L_{3/2}/L_\nu=(R_{3/2}/R_\nu)^2[T(R_{3/2})/T(R_\nu)]^4
\end{equation}
where $R_\nu$ is the radius of the neutrinosphere. To proceed further we use the
temperature profile:
\begin{equation} \label{temp}
T=T_c(R_c/r)^{m/3}
\end{equation}
which is a consequence of eq.(22) and the assumption of local thermal equilibrium. Now,
taking $m=7$\cite {mogri}, we obtain
\begin{equation} \label{SB2}
L_{3/2}/L_\nu=(R_\nu/R_c)^{22/3}[(16Y_e/\pi\alpha^2)(\Lambda^2/M)^4/\rho_cR_cm_p]^{11/9}
\end{equation}
 By demanding that $L_{3/2}\leq0.1L_\nu$ and using $R_\nu\simeq30Km$, we get
\begin{equation} \label{upbnd}
\Lambda\leq 70GeV.
\end{equation}
This in turn implies $m_{3/2}\leq10^{-6}eV$. Since, on the other hand, the
anomalous magnetic moment of the muon already requires $m_{3/2}$ to be larger
than
$\sim10^{-6}eV$ \cite{men,ferrer}, we are forced to conclude that 
\begin{equation} \label{lobnd}
\Lambda\geq300GeV
\end{equation}
or, equivalently,
\begin{equation} \label{last}
m_{3/2}\geq2.2\times10^{-6}eV.
\end{equation}

 In conclusion, we have carefully derived the bounds on the superlight
gravitino mass (i.e. the SUSY scale $\Lambda$) that follow from SN physics.
These limits are completely general in the sense that they do not rely on
other particles in a given particular model  being light.

 Should other
particles such as the scalar partners of the goldstino also be light, then the
resulting bounds are necessarily tighter. In such clearly less general frame,
constraints have also been derived in the literature. Before reviewing these
bounds, let us refer to the very recent claim by Clark et al.\cite{clark} that the
dimension 6 operator $\gamma\tilde{G} \tilde{G}$ in $L_{eff}$ derived by
Luty and Ponton is actually not there. If true, the bounds 
derived above are 1 to 2 orders of magnitude worse\cite{brignole}.

 Suppose now that the gravitino and S and/or P are very light ($ \ll T
$). Because now these scalars can be emitted by astrophysical bodies,
one has to consider additional energy-loss channels\cite{nowa,grif}.The relevant
interaction is given by
\begin{equation}\label{primakoff}
e^{-1} L_{int}=-{\kappa\over 4}\sqrt{2\over 3}({m_{\tilde\gamma}\over
m_{3/2}}) (SF^{\mu\nu}F_{\mu\nu}+P\tilde F_{\mu\nu}F^{\mu\nu})
\end{equation}

where $\kappa \equiv (8\pi)^{1/2}M_{Pl}^{-1}$.

 The main energy drain mechanism is the Primakoff process $ \gamma e
\rightarrow e S/P $ via one photon exchange in the t channel. Calculation
of luminosities in the supernova goes along the same lines as before using, of
course, the Primakoff cross section and which was derived in reference\cite{nowa}.The
restriction 
\begin{equation}\label{lumagain}
L\leq10^{52}erg
\end{equation}     
then implies
\begin{equation}\label{bound1}
m_{3/2}\geq 30({ m_{\tilde \gamma} \over {100GeV}
}) eV
\end {equation}

 Since 
\begin{equation}\label{width}
\Gamma (S/P\rightarrow 2\gamma
)\simeq{\kappa^{2}\over{96\pi}}({m_{\tilde\gamma}\over m_{3/2}})^{2}m_{S/P}^{3}
\end{equation}
sufficiently heavy S/P will decay inside the core. For this NOT to happen, it
is required that
\begin{equation}\label{upbnd}
m_{S/P}\leq 10 MeV
\end{equation}

 Also, S/P leave the supernova core without further rescattering provided
\begin{equation}\label{lbnd}
m_{3/2}\geq 0.3 eV
\end{equation}

(because their m.f.p. is $\sim10^{10}({m_{3/2}\over{30 eV}})^{2}cm$).

For masses below about 0.3 eV, S/P get trapped and their energy is radiated
from a S/P-sphere following the Stefan-Boltzmann law $ L\propto
R_{S/P}^{2}T_{S/P}^{4}$ . The luminosity L is compatible again with
observation (SN1987A) for $m_{3/2}\leq 10^{-1.5} eV$.

{\it Limits from the Sun}\cite{grif}

 Should $m_{S/P}< 1 keV$ then these particles could be emitted from the sun
and the  lower limits to their masses are in this case slightly different. In
the sun, the m.f.p. of S/P is
\begin{equation}\label{mfp}
\lambda_{S/P}\sim 10^{41}({m_{3/2}\over m_{\tilde\gamma}})^{2} cm
\end{equation}

$\lambda_{S/P}$ exceeds the solar radius for $m_{3/2}\geq 10^{-3.5}eV$

(for $m_{\tilde\gamma}=100 GeV$). The
emissivity via Primakoff scattering turns out to be 
\begin{equation}\label{emiss}
\kappa^{2}\alpha({m_{\tilde\gamma} \over
m_{3/2}})^{2}T^{7}F(\kappa_{D}/T)V_{Sun}
\end{equation}

where the function $F(x)$ (see \cite{raffelt}, p. 169) takes care of plasma effects
characterized by the Debye momentum $ \kappa_{D}$. This emissivity is bounded above by
the sun luminosity, i.e. $10^{33} erg/s$. As a consequence, the gravitino mass should
verify
\begin{equation}\label{50eV}
m_{3/2}\geq 50 eV
\end{equation}

 On the other hand, in the trapping regime (i.e. $m_{3/2}<10^{-3.5}$), S/P
emission is allowed as long as $m_{3/2}\leq 10^{-6}$ for in this case their
thermal radiation is sufficiently slowed down. Since $(g-2)_{\mu}$ implies
already $m_{3/2}\geq 10^{-6}$\cite{ferrer} we conclude
\begin{equation}\label{A}
m_{3/2}\geq 50 eV
\end{equation}

for $m_{3/2}<1 keV$ and for $m_{\tilde\gamma}\sim {\it O}(100)GeV$.

{\bf 2b. Big Bang Nucleosynthesis}

 Primordial Helium-4 abundance depends on the effective d.o.f. at
nucleosynthesis because $H\sim g_{*}^{1/2}M_{Pl}^{-1}T^2$ and, the larger H
the sooner do weak interactions decouple. Hence, the neutron to proton ratio
is larger and the helium yield rises accordingly.

 The effective number of d.o.f. for a given species i is 
\begin{equation}\label{dof}
g^{i}_{*}(_{bosons}^{fermions})=( _{7/8}^{1})g_{i}({T_{i}\over T})^4        
\end{equation}

 Bounds from $Y_{P}$ are usually presented as  bounds on $\Delta N_{\nu}$
(i.e. number of extra equivalent neutrino species). It is explicitly given by
the formula
\begin{equation}\label{B}
\Delta N_{\nu}(_{boson}^{fermion})=(_{1}^{8/7}){g_{i}\over
2}[{g_{*}(T_{\nu})\over g_{*}(T_{D_{i}})}]^{4/3}
\end{equation}

where $T_{D_{i}}$ is the decoupling temperature of species i.

The latest analysis from the group in Chicago\cite{copi} sets the limit 
\begin{equation}\label{Chi}
\Delta N_{\nu}\leq 1
\end{equation}

which implies NO bound for the superlight gravitino (equivalent to one
neutrino species). That is, the gravitino does not have to decouple from the
cosmological thermal bath prior to the BBN era. On the other hand, if the
particles S/P are light( $m_{S/P}<1 MeV$, i.e. relativistic at
nucleosynthesis) then, they should freeze out before the Universe cools down
to $T\sim 200 MeV$ so that $\Delta N_{\nu}$ is very small( i.e.
$g_{*}(T_{\nu})/g_{*}(T_{D{i}}) \ll 1 $).

S/P are kept in equilibrium through $\gamma e^{-}\leftrightarrow S/P e^{-}$
and this rate falls below the Hubble rate H when
\begin{equation}\label{C}
\kappa^{2}\alpha({m_{\tilde\gamma}\over m_{3/2}})^{2}T^{3}\sim
g_{*}^{1/2}{T^{2}\over M_{Pl}}
\end{equation}

Taking $T\sim 200 MeV$ ( and $m_{\tilde\gamma}=100GeV$) we get\cite{grif} 
\begin{equation}\label{D}
m_{3/2}\geq 1 eV.
\end{equation}

{\bf 3. SUMMARY}

 In a wide class of supergravity models with a supersymmetry breaking scale in
the TeV range, the gravitino can be very light. In fact, its mass could lie
anywhere between $1 \mu eV$ and $1 keV$. It is also a generic feature of some
of the recently considered models that the superlight gravitino is accompanied
by a superlight scalar S and pseudoscalar P particles, which are the neutral
scalar partners of the goldstino. For energy scales such that $E\gg m_{3/2}$
the longitudinal component of the gravitino dominates and the gravitino
effectively behaves as a spin-1/2 Goldstino. The neutral scalars and the
gravitino are coupled to matter with strength inversely proportional to the
gravitino mass and, hence, they can be abundantly produced in the interior of
stellar cores. By requiring that the radiated power does not overcome
$10^{52}erg/s$ in a supernova explosion and $10^{33} erg/s$  in the case of
solar emission, we obtain the following results.

\begin{equation}\label{E}
m_{3/2}\geq 10^{-5} eV
\end{equation}
independently of $m_{3/2}$. Also,
\begin{equation}\label{F}
m_{3/2}\geq 30 eV  
\end{equation}
 or
\begin{equation}\label{G}
m_{3/2}\leq 0.03 eV
\end{equation}

if $m_{S/P}\leq 10 MeV$ and,
\begin{equation}\label{H}
m_{3/2}\geq 50 eV
\end{equation}
for $m_{S/P}\leq 1 keV$.

 Finally, from Big Bang Nucleosynthesis arguments  we infer the limit
\begin{equation}\label{I}
m_{3/2}\geq 1 eV
\end{equation}

should  $m_{S/P}$ be lighter than about 1 MeV.

 {\it I am grateful to the organizers of the Workshop for the friendly hospitality  and
for the nice atmosphere provided in Pune.

 Work partially supported by the CICYT Research Project 
AEN95-0882}.

\end{document}